\documentclass[12pt,twocolumn]{iopart}

\usepackage{iopams} 
\usepackage{graphicx} 
\begin{document}

\title[Thermal conductivity $\beta$-$\mathrm{Ga_2O_3}$]{Temperature-dependent thermal conductivity in Mg-doped and undoped $\beta$-$\mathrm{Ga_2O_3}$ bulk-crystals}

\author{M. Handwerg$^{1,2}$, R. Mitdank$^{1}$, Z. Galazka$^{3}$, S. F. Fischer$^{1}$}

\address{1. AG Neue Materialien, Humboldt-Universit{\"a}t zu Berlin, 10099 Berlin, Germany} \address{2. Helmholtz-Zentrum Berlin f{\"u}r Materialien und Energie GmbH, Hahn-Meitner-Platz 1, 14109 Berlin, Germany} \address{3. Leibniz Institute for Crystal Growth, Max-Born-Strasse 2, 12489 Berlin, Germany}
\ead{handwerg@physik.hu-berlin.de}

\begin{abstract}
For $\beta$-$\mathrm{Ga_2O_3}$ only little information exist concerning the thermal properties, especially the thermal conductivity $\lambda$. 
Here, the thermal conductivity is measured by applying the electrical 3$\omega$-method on Czochralski-grown $\beta$-$\mathrm{Ga_2O_3}$ bulk crystals, which have a thickness of $200~\mathrm{\mu m}$ and $800~\mathrm{\mu m}$. At room temperature the thermal conductivity along the [100]-direction in Mg-doped electrical insulating and undoped semiconducting $\beta$-$\mathrm{Ga_2O_3}$ is confirmed as $13\pm 1~\mathrm{Wm^{-1}K^{-1}}$ for both crystals. The thermal conductivity increases for decreasing temperature down to 25~K to $\lambda(25~\mathrm{K})=(5.3\pm 0.6)\cdot 10^2~\mathrm{Wm^{-1}K^{-1}}$. 
The phonon contribution of $\lambda$ dominates over the electron contribution below room temperature. 
The observed function $\lambda(T)$ is in accord with phonon-phonon-Umklapp scattering and the Debye-model for the specific heat at $T\gtrsim 90~\mathrm{K}$ which is about $0.1$ fold of the Debye-temperature $\theta_\mathrm{D}$. Here a detailed discussion of the phonon-phonon-Umklapp scattering for $T< \theta_\mathrm{D}$ is carried out. The influence of point defect scattering is considered for $T<100~\mathrm{K}$.

\end{abstract}

\maketitle

\section{Introduction}
$\beta$-$\mathrm{Ga_2O_3}$ is an important material for high-power electronic applications \cite{introhighpower} because of the wide band gap of about 4.8~eV at room temperature \cite{introbandgap}. Especially, the transparent and semiconducting properties make it useful for applications in opto-electronics, like thin-film electroluminescent displays \cite{transdis} and transparent field effect transistors \cite{transFET}.\\ 
There exists five different modifications of $\mathrm{Ga_2O_3}$ ($\alpha$ - $\epsilon$), but only the monoclinic $\beta$-$\mathrm{Ga_2O_3}$ is the most stable modification. While experimental data about optical and electrical \cite{mitdank,introbulk} properties for bulk material and thin films have been investigated, rare information exists concerning the thermal properties, especially their temperature dependence. The low-temperature dependence of the thermal conductivity gives information about the scattering mechanisms and the dominating heat transfer process.
In general, thermal properties are a prerequisite for application in high-power electronics, because heat transfer processes play an important role.
At room temperature, measurements of the thermal conductivity were performed by the laser flash method and revealed a value of $\lambda_\mathrm{ref}=13~\mathrm{Wm^{-1}K^{-1}}$ in the [100]-direction \cite{Referenz1} and $\lambda_\mathrm{ref}=21~\mathrm{Wm^{-1}K^{-1}}$ along the [010]-direction \cite{Galazka2}.
Below room temperature, the thermal conductivity has not been reported to date.\\
Here, we investigated the thermal conductivity of undoped semiconducting n-type $\beta$-$\mathrm{Ga_2O_3}$ and Mg-doped insulating $\beta$-$\mathrm{Ga_2O_3}$ for a wide temperature range from 25~K to 301~K.
Low doping concentrations of Mg in $\beta$-$\mathrm{Ga_2O_3}$ invoke complete electrical insulating behavior, which is important to determine the phonon part of the thermal conductivity, because free electrons can carry heat as well.
The thermal conductivity was determined by the so called 3$\omega$-method \cite{Cahill1}.

\section{Sample preparation}

\begin{figure}[h]
\includegraphics[width=0.75\columnwidth,keepaspectratio]{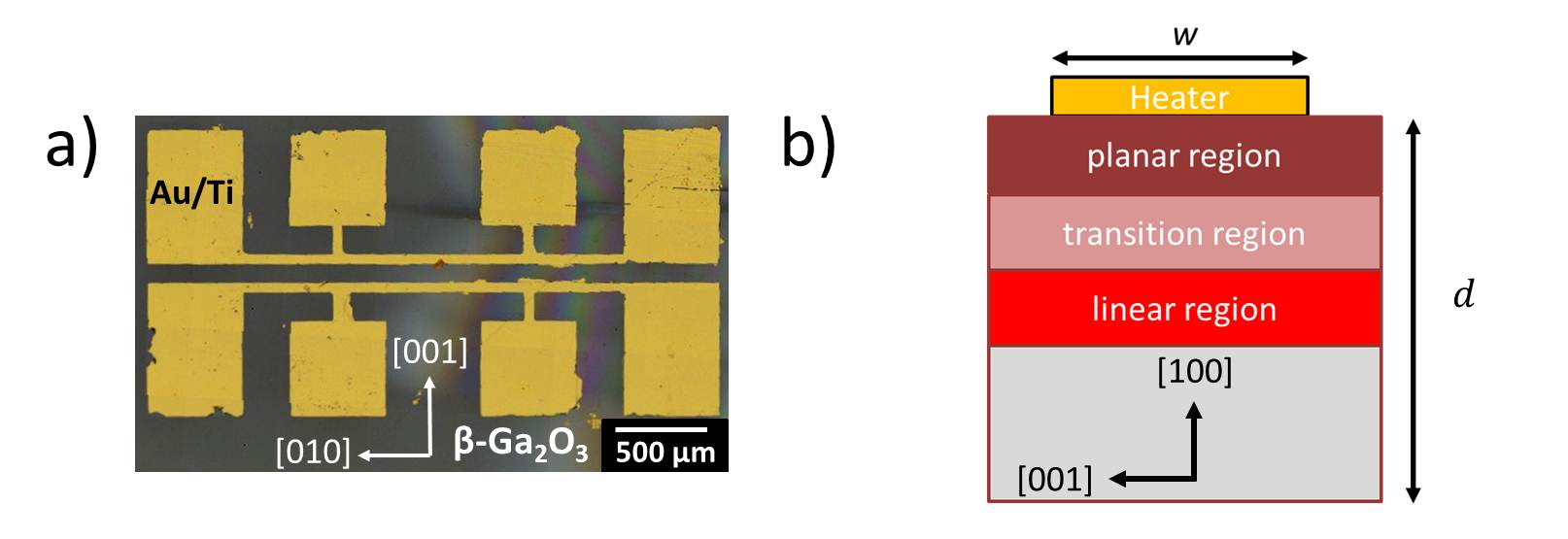}
\caption{a) An arrangement of two line heater lines on top of the Mg-doped $\beta$-$\mathrm{Ga_2O_3}$-crystal. Every line heater can be used to determine the thermal conductivity in the direction normal to the surface, here [100]. b) A schematic cross-section of the crystal with a thickness $d=800~\mathrm{\mu m}$ and a heater width $w=50~\mathrm{\mu m}$, which is separated in 3 regions (planar, transition, linear). Only in the linear region the heat flow is homogeneous which is a prerequisite for the determination of the thermal conductivity from the measurement. The heater parameters are listed in table \ref{table}.}
\label{microbild}
\end{figure}
$ $\\
Bulk $\beta$-$\mathrm{Ga_2O_3}$ single crystals were obtained from the melt by the Czochralski method with use of an iridium crucible and a dynamic, self-adjusting growth atmosphere to minimize decomposition of $\mathrm{Ga_2O_3}$ and oxidation of iridium crucible, as described in detail in ref. \cite{Galazka1,Galazka2}. Semiconducting (undoped) and insulating (doped with Mg) crystals of 22 mm diameter were grown along [010]-direction. From bulk single crystals double side epi-polished wafers of 0.5x5x5$~\mathrm{cm^3}$ were prepared. Both semiconducting and insulating wafers were [100]-oriented. Semiconducting wafers had a free electron concentration of $n=7\cdot 10^{17}~\mathrm{cm^{-3}}$, with a resistivity of $\rho= 0.1~\Omega \mathrm{cm}$ and an electron mobility of $\mu= 100~\mathrm{cm^{2} V^{-1} s^{-1}}$ as obtained from Hall effect measurements in the van-der-Pauw configuration with use of In-Ga ohmic contacts.
In order to invoke insulating behavior the electron concentration was compensated by an acceptor concentration of $n_\mathrm{Mg}=11~\mathrm{wt.ppm. Mg}$. 
The mean impurity distance can be estimated to $x=16~\mathrm{nm}$.\\ 
The heater is patterned with laser lithography using the positive resist AZECI 3027. A titanium layer of about 10~nm and a conducting gold layer of about 50~nm were sputtered. 
Subsequently, a lift-off was performed with acetone in an ultrasonic bath.
Figure \ref{microbild} shows two typical line heater structures. Finally, the sample is mounted in a chip-carrier and bonded with Al wires.

\section{Thermal Conductivity Measurement} 

\begin{figure}[h]
\includegraphics[width=0.40\columnwidth,keepaspectratio]{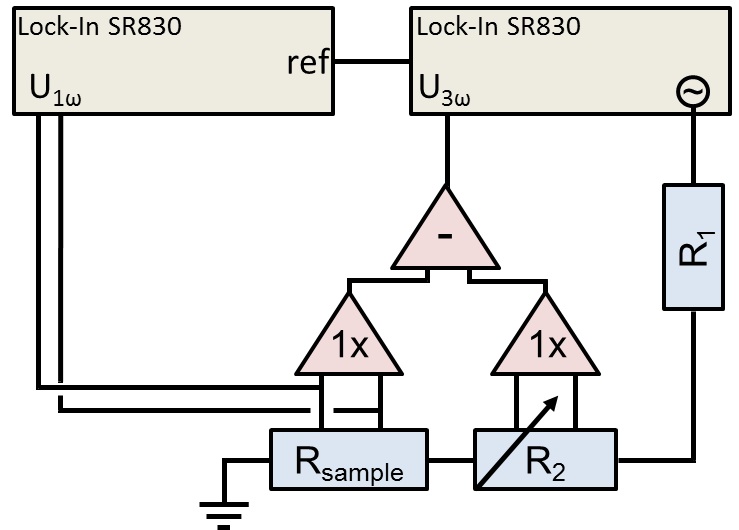}
\caption{Measurement setup with two Lock-In amplifiers SR830 and a differential amplifier to measure the voltage signals $U_{1\omega}$ and $U_{3\omega}$. As voltage source the internal function generator is used.}
\label{Plan}
\end{figure}

The 3$\omega$-method, first implemented by David Cahill \cite{Cahill2} is based on the resistance increase due to an alternating current $I$ passing through an electrical heater line.
A conducting (Au) metal strip of the resistance $R$ is placed on top of the investigated sample and produces heat with the power of $P=I^2\cdot R$.
With increasing temperature $T=T_0+\Delta T$ the resistance $R$ increases as $R(\Delta T)=R_0\cdot(1+\alpha_0\Delta T)$, where $\alpha_0$ denotes the temperature coefficient $\alpha_0=(1/R_0) \cdot \mathrm{d}R/\mathrm{d}T$ and $R_0$ the resistance at bath temperature $T_0$.\\
Because the sample is heated with an alternating current $I$, the depending parameters such as the power $P$, the temperature change $\Delta T$ and the resistance $R$ alternate as well but with the double angular frequency $2\omega$ which is related to the frequency $f$ by $\omega=2 \pi f$.
Therefore the measured AC-voltage is $U=I(\omega)\cdot R(2\omega)=U_{1\omega}+U_{3\omega}$, where $U_{1\omega}$ and $U_{3\omega}$ are voltage contributions that oscillate with the frequencies $1\omega$ and $3\omega$, respectively. The contribution $U_{3\omega}$ yields information about the temperature change of the heater line \cite{Cahill2} 
\begin{equation}\Delta T=\frac{2}{\alpha_0}\cdot \frac{U_{3\omega}}{U_{1\omega}} = \frac{P}{\pi l \lambda}\left(\frac{1}{2}\ln\left(\frac{D}{r^2}\right)+\frac{1}{2}\ln(2\omega)-\frac{i\pi}{4}-\gamma\right)
\end{equation}
depending on the thermal conductivity $\lambda$, the thermal diffusivity $D$, the distance $r$ and the Euler gamma constant $\gamma$.   
The thermal conductivity of the sample is given by 
\begin{equation}
\lambda=\frac{1}{4\pi l}\left[\frac{\Delta\ln(2\omega)}{\Delta U_{3\omega}}\right]\frac{U_{1\omega}^3}{R^2}\frac{\mathrm{d}R}{\mathrm{d}T}.
\end{equation}
Here, the ratio $\Delta U_{3\omega}/\Delta\ln(2\omega)$ represents the experimentally determined slope of the plot $U_{3\omega}(\ln(f))$ and $l$ the distance between the inner voltage contacts. 
The heating current is induced in the outer contacts of the measurement structure, whereas both voltages  $U_{1\omega}$ and $U_{3\omega}$ were measured at the inner contacts by a Lock-In amplifier SR830.
In order to be able to measure the voltage $U_{3\omega}$ that is superimposed by a voltage of $U_{1\omega}$, the voltage signal of $U_{1\omega}$ is compensated using a differential amplifier and a variable resistor $R_2$. The measurement setup is depicted in figure \ref{Plan}. All measurements were performed in high vacuum, so that thermal convection cannot influence the results. Thermal radiation plays a minor role in the used temperature range \cite{volklein}.\\
In order to perform this measurement, a well defined heater line is required.
Especially the heater width has to compare with the thermal penetration depth $q^{-1}=\sqrt{\frac{D}{2\omega}}$ with the thermal diffusivity $D=\lambda/(c_p\rho_m)$, the specific heat $c_p$ and the density $\rho_m$.
To get a homogeneous heat flow into the sample, the thermal penetration depth must be at least the heater width. Furthermore, the maximal penetration depth $q^{-1}$ should not exceed the crystal thickness $d$ to avoid the penetration of heat into the range outside of the sample.
Within these limits $U_\mathrm{3\omega}(\ln f)$ is linear \cite{eindringtiefe}.\\
We estimated the valid frequency range by using the reference value $\lambda_\mathrm{ref}=13~\mathrm{Wm^{-1}K^{-1}}$ \cite{Referenz1}. 
With a crystal thickness $d$ of $200~\mu\mathrm{m}$ (semiconducting $\beta$-$\mathrm{Ga_2O_3}$) and  $800~\mu\mathrm{m}$ (insulating $\beta$-$\mathrm{Ga_2O_3}$) and the heater parameters from table \ref{table} the frequency range can be estimated to $f_{200~\mu\mathrm{m}}\approx 50~\mathrm{Hz}-250~\mathrm{Hz}$ and $f_{800~\mu\mathrm{m}}\approx 2~\mathrm{Hz}-150~\mathrm{Hz}$, respectively.
Because the thermal conductivity increases with decreasing $T$, the limits shift to higher frequencies. The measurement setup has a low-pass character with a cutoff frequency that influences $U_\mathrm{1\omega}$ and gives an upper frequency limit. Especially for thinner layers the frequency range is reduced. Due to this fact, measurements of the 200~$\mu$m sample could only be carried out for $T>170~\mathrm{K}$. 

\section{Experimental Results} 
\begin{figure}[h]
\includegraphics[width=0.80\columnwidth,keepaspectratio]{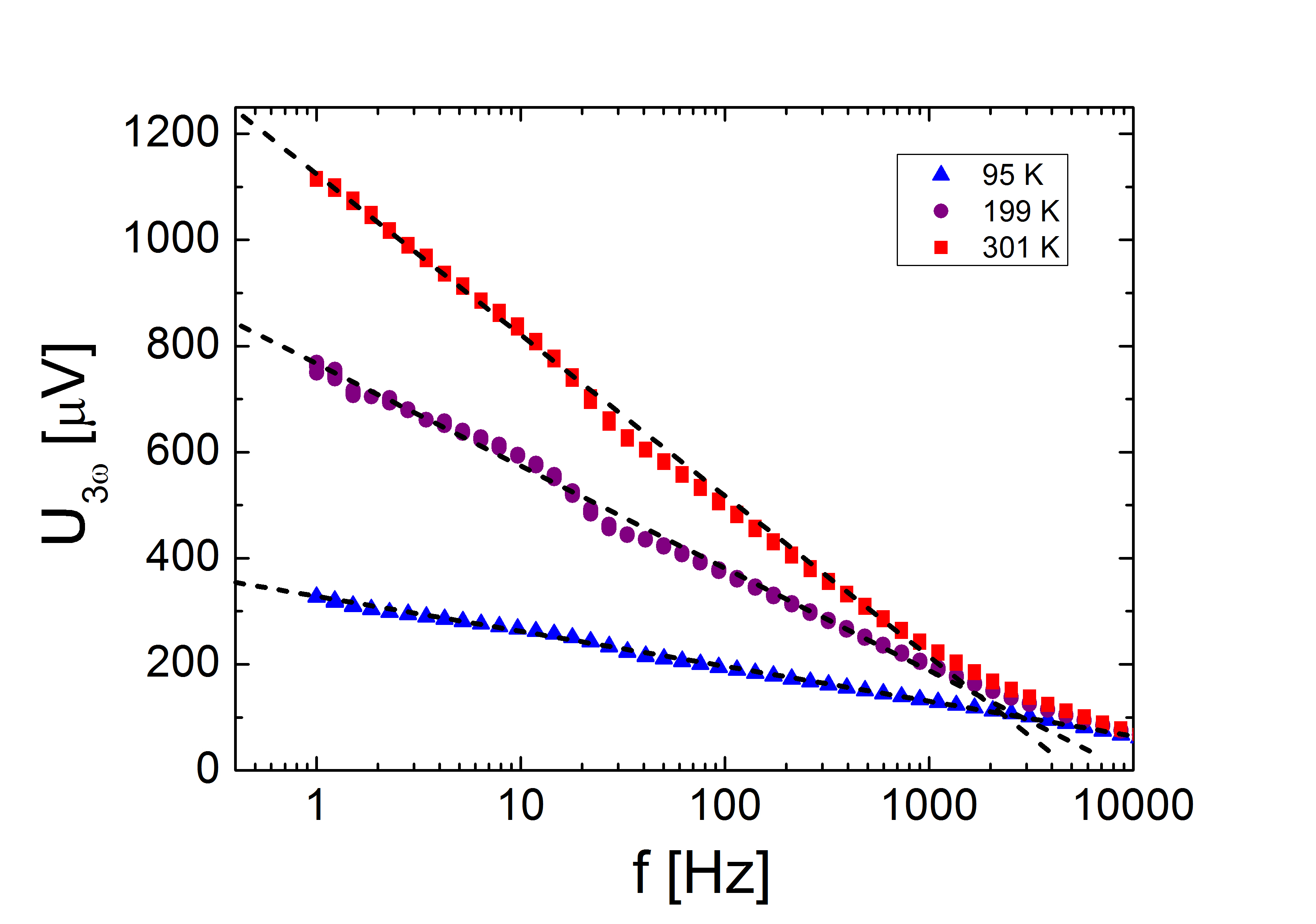}
\caption{Frequency-dependent voltage $U_{3\omega}$ with a linear approximation in a semilogarithmic plot, measured with a heater line on top of the Mg-doped $\beta$-$\mathrm{Ga_2O_3}$ crystal. This is an exemplary plot for several temperatures for one heater and one crystal.}
\label{U3(f)}
\end{figure}

The measurements for the thermal conductivity were performed separately for two heater lines on insulating Mg-doped $\beta$-$\mathrm{Ga_2O_3}$ as seen in figure \ref{microbild} and one heater line on undoped semiconducting $\beta$-$\mathrm{Ga_2O_3}$.
For several temperatures the $U_{3\omega}$ was measured and is shown exemplary in figure \ref{U3(f)}. The voltage $U_{1\omega}$ was measured and remained frequency-independent as expected.\\

$ $\\
The $U_{3\omega}$-voltage is a linear function of $\ln f$ in the frequency ranges estimated for our crystal thicknesses. Therefore, equation 2 can be used to determine the thermal conductivity $\lambda$.

\begin{figure}[h]
\includegraphics[width=0.80\columnwidth,keepaspectratio]{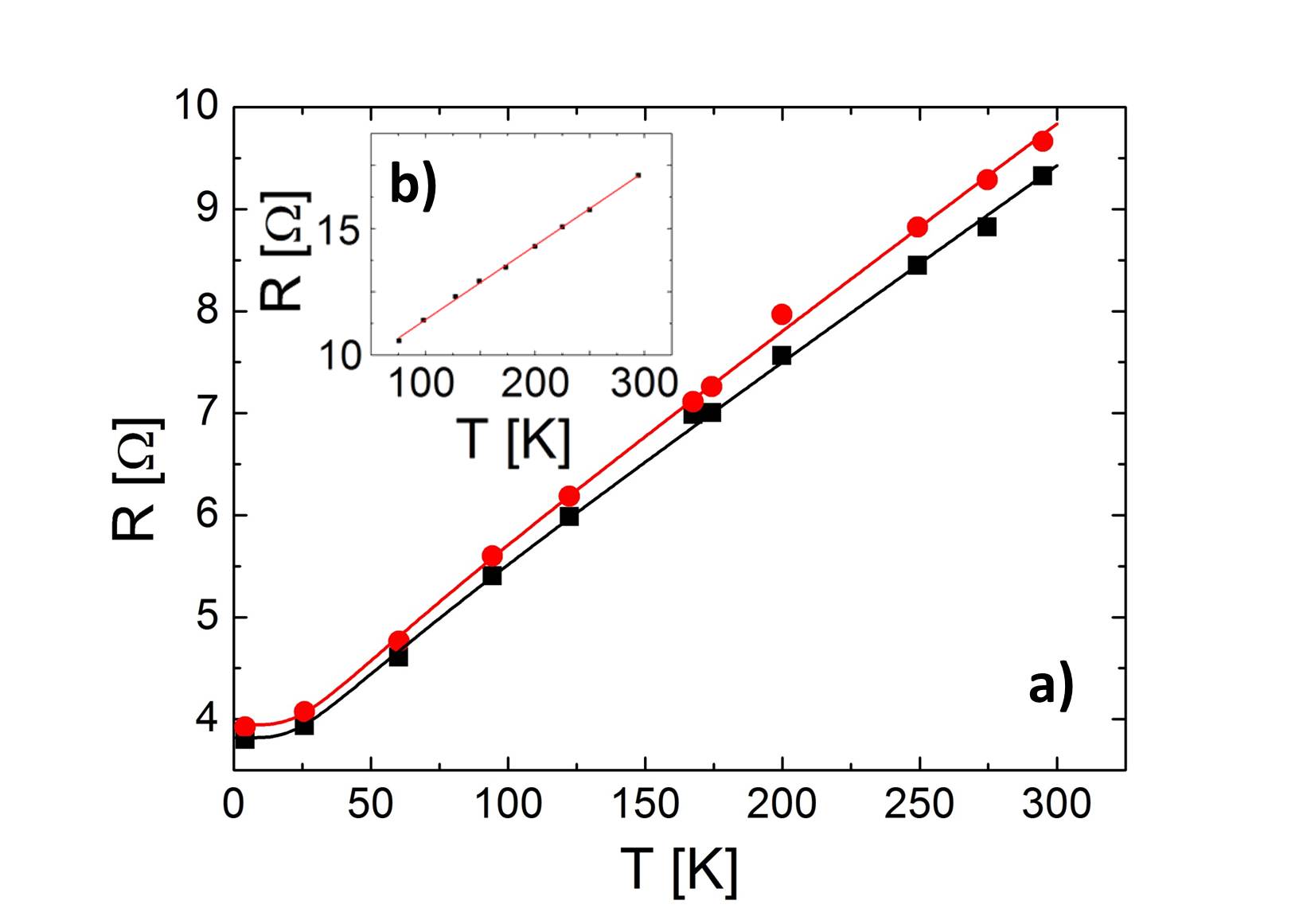}
\caption{Temperature-dependent resistance of the three heater lines on insulating (a) and semiconducting (b) $\beta$-$\mathrm{Ga_2O_3}$. The slope of the Bloch-Gr{\"u}neisen-approximations and the resistance values are used to determine the thermal conductivity with equation 2.}
\label{R(T)}
\end{figure}
\begin{figure}[h]
\includegraphics[width=0.75\columnwidth,keepaspectratio]{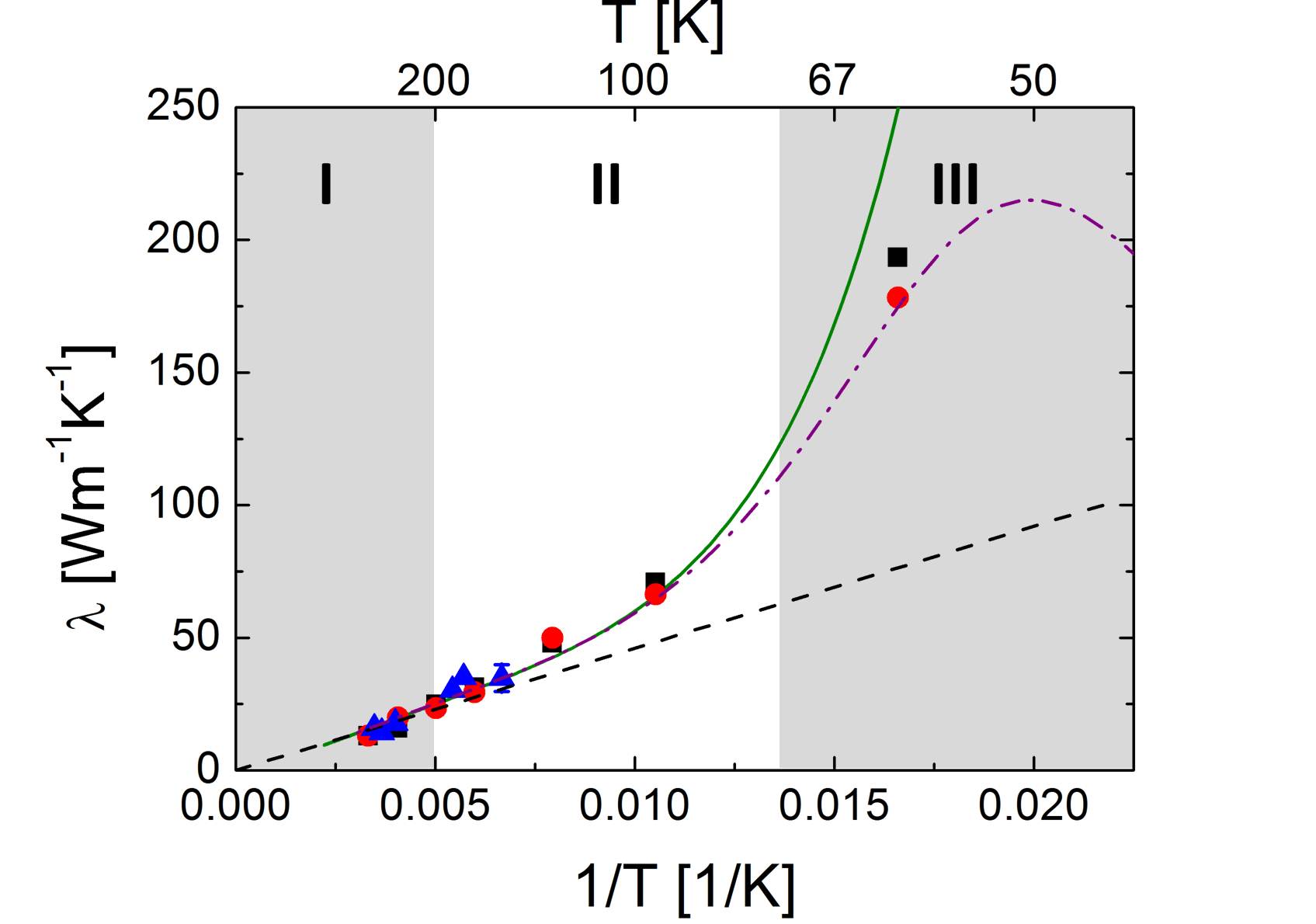}
\caption{The temperature dependent thermal conductivity of Mg-doped $\mathrm{\beta-Ga_2O_3}$ (black squares, red circles) and undoped $\mathrm{\beta-Ga_2O_3}$ (blue triangles). Dashed line: phonon-phonon scattering for $T\gg\theta_\mathrm{D}$. Solid line: phonon-phonon-Umklapp scattering, calculated for $T<\theta_\mathrm{D}$. Dot-dashed line: additional scattering process using Matthiessens rule for a temperature independent $\Lambda_\mathrm{imp}=1.5~\mathrm{\mu m}$. The regions for high (I), intermediate (II) and low temperature (III) are explained in the discussion.}
\label{lambdat}
\end{figure}

$ $\\
As seen in the equation 2 it is important to determine the heater characteristics including the length $l$, the width $w$ , the resistance $R_0$, the temperature coefficient $\alpha_0$ and the variation of the resistance with temperature $\mathrm{d}R/\mathrm{d}T$. These values are shown in table \ref{table}. The temperature-dependent resistance of the sputtered Au/Ti-heater line is plotted in figure \ref{R(T)}. The function $R(T)$ satisfies the Bloch- Gr{\"u}neisen-Law \cite{tritt}
\begin{equation}
R(T)-R(4.2~\mathrm{K})\propto \left(\frac{T}{\theta_\mathrm{D}}\right)^5\int_0^{\theta_\mathrm{D}/T}\frac{x^5 }{(e^x-1)(1-e^{-x})}\mathrm{d}x
\end{equation}
with an approximated Debye-temperature $\theta_\mathrm{D}$ of $145(5)~K$. The Debye-temperature of pure Gold is $165~K$ \cite{kittel}, but we expected a lower value for our sputtered material. The slope $\mathrm{d}R/\mathrm{d}T$ is given by the derivation.
\begin{table}
\begin{tabular}{ c | c | c | c  }
sample & $\beta$-$\mathrm{Ga_2O_3}$ & Mg-doped $\beta$-$\mathrm{Ga_2O_3}$ (heater 1) & Mg-doped $\beta$-$\mathrm{Ga_2O_3}$ (heater 2)\\
\hline
$l$ & $1.21(4)~\mathrm{mm}$ & $1.00(5)~\mathrm{mm}$ & $1.00(5)~\mathrm{mm}$\\
$w$                  & $40(1)~\mathrm{\mu m}$& $50(1)~\mathrm{\mu m}$& $50(1)~\mathrm{\mu m}$\\
$R_0$			     & $16.4(1)~\Omega$      & $9.3(1)~\Omega$ & $9.7(1)~\Omega$\\
$\alpha_0$ 			 & $0.00185(4) ~\mathrm{K^{-1}} $ & $0.00206(4) ~\mathrm{K^{-1}}$ & $0.00210(4) ~\mathrm{K^{-1}}$\\
$\mathrm{d}R/\mathrm{d}T$ & $0.0303(6)~\Omega \mathrm{K^{-1}}$ & $0.0191(4)~\Omega \mathrm{K^{-1}}$ & $0.0204(4)~\Omega \mathrm{K^{-1}}$
\end{tabular}
\caption{Heater parameter for the 3 heater lines used in this paper. $l$ is the distance of the voltage contacts to measure $U_{1\omega}$ and $U_{3\omega}$ in 4-point geometry. $R_0$ and $\alpha_0$ are specified for RT. The values in the brackets denote the uncertainty in the last digit.}
\label{table}
\end{table}
$ $\\
In figure \ref{lambdat}, the measured thermal conductivity values are shown as a function of the inverse temperature in the interval from $60~\mathrm{K}<T<301~\mathrm{K}$ for both, the insulating and the semiconducting $\beta$-$\mathrm{Ga_2O_3}$ crystal.
An increase of the thermal conductivity with decreasing temperature is observed. Both, the insulating Mg-doped and the undoped semiconducting $\beta$-$\mathrm{Ga_2O_3}$ crystals have the same thermal conductivity values within the measurement uncertainty.
The room temperature value of the thermal conductivity is $\lambda(301~\mathrm{K})=13\pm 1~\mathrm{Wm^{-1}K^{-1}}$. 
The accuracy of the thermal conductivity according to equation 2 is mainly determined by the measurement uncertainties of $l$ (see fig.\ref{microbild}), $\mathrm{d}R/\mathrm{d}T$ (see fig.\ref{R(T)}) and the slope of the function $U_\mathrm{3\omega}(\ln f)$ (see fig.\ref{U3(f)}). The theoretically expected temperature dependence seen in figure \ref{lambdat} is discussed below. 

\begin{figure}[h]
\includegraphics[width=0.75\columnwidth,keepaspectratio]{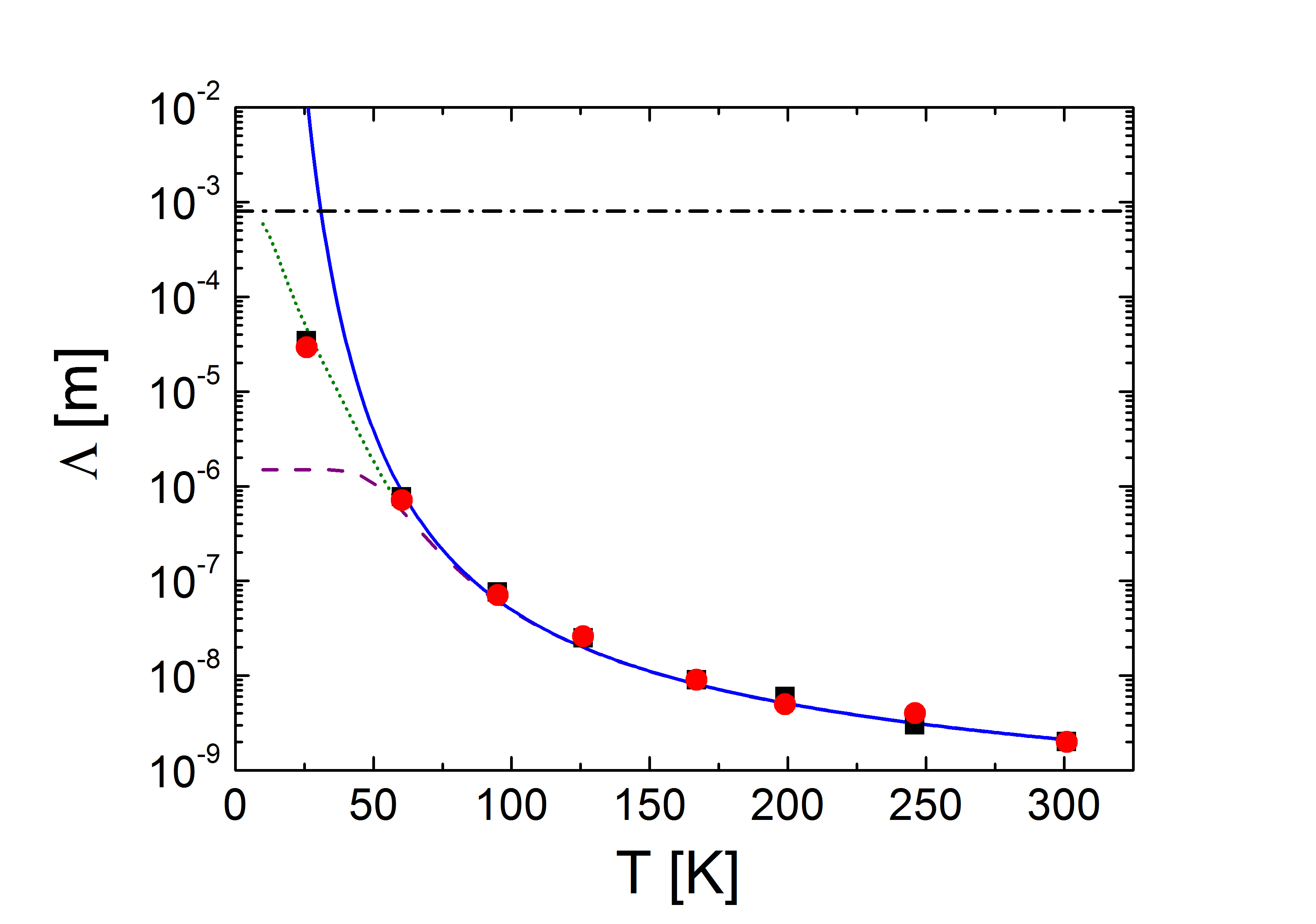}
\caption{The temperature dependent mean free path of phonons in the insulating Mg-doped $\mathrm{\beta-Ga_2O_3}$, calculated from the measured thermal conductivities. The solid line is the theoretical contribution for phonon-phonon-Umklapp-scattering and the dashed line shows additionally a contribution of a second scattering process with a constant mean free path of $1.5~\mathrm{\mu m}$. The dotted line describes point-defect-scattering within the scope of Rayleigh scattering.}
\label{mfp}
\end{figure}
$ $\\

\section{Discussion and outlook} 
The thermal conductivity $\lambda$ has an electron and a phonon contribution $\lambda=\lambda_\mathrm{el}+\lambda_\mathrm{ph}$. Within the experimental error, both semiconducting and insulating $\mathrm{\beta-Ga_2O_3}$ crystals show the same values and temperature dependence of the thermal conductivity $\lambda(T)$. Therefore, we conclude that an electronic contribution to the thermal conductivity can be considered minor and the phonon contribution dominates. To ensure, that only the phonon contribution is relevant for our discussion, the free electron contribution to the thermal conductivity can be estimated at $T=300~\mathrm{K}$ using the Wiedemann-Franz-law $\lambda_\mathrm{el}=L_0 \sigma T$ with the Lorenznumber $L_0=2.44\cdot 10^{-8}~ \mathrm{W\Omega K^{-2}}$. The resistivity was determined as $\rho=\sigma^{-1}=0.1~\mathrm{\Omega cm}$ for the investigated semiconducting $\beta$-$\mathrm{Ga_2O_3}$-crystal. Hence, the free electron contribution for the thermal conductivity is $\lambda_\mathrm{el}=7\cdot10^{-3}~\mathrm{Wm^{-1}K^{-1}}$, which is 3 orders of magnitude lower than our measured values. Therefore, we conclude, that below room temperature relevant heat is only carried by phonons.\\
In figure \ref{lambdat} the measured thermal conductivity $\lambda$ is plotted as a function of the reciprocal temperature between 60~K and room temperature (RT). We discuss three regions. Region I corresponds to the high-temperature behavior. As a result, a relationship $\lambda = A/T$  was found for $T>200~\mathrm{K}$ with a constant $A$ which can be approximated by $A \approx 4600~\mathrm{Wm^{-1}}$. An inverse dependence of the thermal conductivity on the temperature $\lambda\propto T^{-1}$ is expected in insulators in the case of Umklapp scattering if the high temperature condition $T >> \theta_\mathrm{D}$ (Debye temperature) is fulfilled.
However, in our measurement the bath temperature lies systematically below the Debye temperature $T < \theta_\mathrm{D}$ with $\theta_\mathrm{D} = 870~\mathrm{K}$ \cite{Haying}.\\
In region II, the measured values start to differ from the expected inverse behavior and a more specific examination of the thermal conductivity should be carried out. Therefore, we calculate the thermal conductivity as $\lambda(T)=(1/3)\Lambda(T)\cdot c_V(T)\cdot v_s$ with the mean free path $\Lambda$, the specific heat $c_V$ and the velocity of sound $v_s$, using the Debye model for specific heat and phonon-phonon-scattering as the dominating scattering process.\\
The specific heat is given by \cite{tritt}
\begin{equation}
c_V\propto\left(\frac{T}{\theta_\mathrm{D}}\right)^3\int_0^{\theta_\mathrm{D}/T}\frac{x^4 e^x}{(e^x-1)^2}\mathrm{d}x
\end{equation} 
and the mean free path is given by
\begin{equation}
\Lambda_\mathrm{ph-ph}\propto \left(e^{\theta_\mathrm{D}/2T}-1\right) \quad.
\end{equation} 
As seen in figure \ref{lambdat}, the calculated $\lambda(T)$ agrees well with our experimental values. Therefrom we conclude, that in compensated insulating Mg-doped $\beta$-$\mathrm{Ga_2O_3}$ for $T \gtrsim 100~\mathrm{K}$ the phonon-phonon-Umklapp scattering determines the mean free path $\Lambda$. A similar  temperature dependence is also found for Si with $\theta_\mathrm{D} \approx 630~\mathrm{K}$. There, an increase of $\lambda$ with decreasing temperature is observed for $T/\theta_\mathrm{D} > 0.1$ \cite{silicon}.\\
In figure \ref{mfp}, the mean free path $\Lambda$ is plotted as a function of $T$. The absolute value of $\Lambda$ was calculated by using the room temperature values of $\lambda = 13~\mathrm{Wm^{-1}K^{-1}}$, the velocity of sound $v_s=6.2\cdot 10^3~\mathrm{ms^{-1}}$ (estimated from Youngs modul) and the specific heat $c_V = 2.9\cdot 10^6~ \mathrm{JK^{-1}m^{-3}}$. The mean free path increases from $\Lambda=2.1~\mathrm{nm}$ at $T=300~\mathrm{K}$ to $\Lambda=0.1~\mathrm{\mu m}$ at $100~\mathrm{K}$.\\
Region III depicts the low temperature behavior of the thermal conductivity.
For $T\leq 100~\mathrm{K}$ we found that $\lambda(T)$ deviates from the approximation for Umklapp-scattering. Therefore, we discuss an additional influence of impurity scattering on the mean free path. 
By applying the Matthies’sens rule
\begin{equation}
\frac{1}{\Lambda}=\frac{1}{\Lambda_\mathrm{ph-ph}}+\frac{1}{\Lambda_\mathrm{imp}}
\end{equation}
we found the typical function $\lambda(T)$ for an insulator. The dashed curve in figure \ref{mfp} was plotted for a temperature independent $\Lambda_\mathrm{imp}=1.5~\mathrm{\mu m}$, which agrees well with the data down to $T=60~\mathrm{K}$.\\
If we assume a temperature dependent point defect scattering with 
\begin{equation}
\Lambda_\mathrm{imp}(T)=\Lambda_\mathrm{imp}(T_0)\cdot\left(\frac{T_0}{T}\right)^4
\end{equation} 
we can describe the function $\Lambda(T)$ for the whole temperature range as plotted in figure \ref{mfp} (dotted line).\\
The influence of boundary scattering depends on the sample thickness $d$. Thin layers with $d < 100~\mathrm{nm}$ at $T\lesssim 100~\mathrm{K}$ would almost achieve the Casimir limit. If $d=\Lambda$, from figure \ref{mfp} the threshold temperature of the boundary scattering can be determinated. The influence of phonon-boundary scattering for $d=800~\mathrm{\mu m}$, is only expected for $T \leq 25~\mathrm{K}$.\\
With respect to future studies, the very low temperature range $T<25~\mathrm{K}$ and the temperature range higher than room temperature remain to be investigated. Furthermore, an anisotropy in the thermal conductivity along the three crystallographic orientations is predicted and needs experimental verification.  

\section{Summary} 
By using the 3$\omega$-method, an increase of the thermal conductivity from $\lambda(301~\mathrm{K})=13\pm 1~\mathrm{Wm^{-1}K^{-1}}$ to $\lambda(25~\mathrm{K})=(5.3\pm 0.6)\cdot 10^2~\mathrm{Wm^{-1}K^{-1}}$ of Mg-doped insulating and undoped semiconducting $\beta$-$\mathrm{Ga_2O_3}$ was measured along the [100]-direction.
The room temperature values of the thermal conductivity was measured as $\lambda=13\pm 1 ~\mathrm{Wm^{-1}K^{-1}}$ and confirms the result obtained by laser flash spectroscopy for the same orientation\cite{Referenz1}.
The temperature dependence of the thermal conductivity was investigated for $T\gtrsim 25~\mathrm{K}$
and can be explained by phonon-phonon-Umklapp scattering. For $T<100~\mathrm{K}$ point defect scattering must be considered.
  
\section{Acknowledgment} 
We like to thank Dr. Srujana Dusari and Christine B{\"u}low for the help with the sample preparation. This work is funded by Deutsche Forschungsgemeinschaft within the framework of DFG SPP1386 Priority Program. M.H. is grateful for support by 'MatSEC Graduate School', Helmholtz-Zentrum Berlin f{\"u}r Materialien und Energie GmbH.
 
\newpage

\section*{References}


\begin{thebibliography}{15}


\providecommand{\url}[1]{\texttt{#1}}
\expandafter\ifx\csname urlstyle\endcsname\relax
  \providecommand{\doi}[1]{doi: #1}\else
  \providecommand{\doi}{doi: \begingroup \urlstyle{rm}\Url}\fi

\bibitem[1]{introhighpower}
\textsc{Grundmann}, M.; \textsc{Frenzel}, H.; \textsc{Lajn}, A.;
  \textsc{Lorenz}, M.; \textsc{Schein}, F.; \textsc{Wenckstern}, H. von:
\newblock {}\emph{Phys. Status Solidi A} 207,  1437--1449, (2010).

\bibitem[2]{introbandgap}
\textsc{Tippins}, H.~H.:
\newblock {}\emph{Phys. Rev.} 140, A316-A319, (1965).

\bibitem[3]{transdis}
\textsc{Miyata}, T.; \textsc{Nakatani}, T.; \textsc{Minami}, T.:
\newblock {}\emph{J. Lumin.} 87–89,  1183–1185 S., (2000).

\bibitem[4]{transFET}
\textsc{Higashiwaki}, M.; \textsc{Sasaki}, K.; \textsc{Kuramata}, A.;
  \textsc{Masui}, T.; \textsc{Yamakoshi}, S.:
\newblock {}\emph{Appl. Phys. Lett.} 100,  013504 S., (2012).

\bibitem[5]{mitdank}
\textsc{Mitdank}, R.; \textsc{Dusari}, S.; \textsc{B{\"u}low}, C.;
  \textsc{Albrecht}, M.; \textsc{Galazka}, Z.; \textsc{Fischer}, S.F.:
\newblock {}\emph{Phys. Status Solidi A} 211,  543--549, (2014).

\bibitem[6]{introbulk}
\textsc{Irmscher}, K.; \textsc{Galazka}, Z.; \textsc{Pietsch}, M.;
  \textsc{Uecker}, R.; \textsc{Fornari}, R.:
\newblock {}\emph{J. Appl. Phys.} 110,  063720 S., (2011).

\bibitem[7]{Referenz1}
\textsc{Villora}, Encarnacion~G.; \textsc{Shimamura}, Kiyoshi; \textsc{Ujiie},
  Takekazu; \textsc{Aoki}, Kazuo:
\newblock {}\emph{Applied Physics Letters} 92,  202118, (2008).

\bibitem[8]{Galazka2}
\textsc{Galazka}, Z.; \textsc{Irmscher}, K.; \textsc{Uecker}, R.; \textsc{Bertram}, R.;
  \textsc{Pietsch}, M.; \textsc{Kwasniewski}, A.; \textsc{Naumann}, M.; \textsc{Schulz}, T.;
  \textsc{Schewski}, R.; \textsc{Klimm}, D.; \textsc{Bickermann}, M.:
\newblock {}\emph{Journal of Crystal Growth} 404, ~184--191, (2014).
\bibitem[9]{Cahill1}
\textsc{Cahill}, David~G.; \textsc{Pohl}, R.~O.:
\newblock {}\emph{Physical Review B} 35,  4067--4073, (1987).

\bibitem[10]{Galazka1}
\textsc{Galazka}, Z.; \textsc{Uecker}, R.; \textsc{Irmscher}, K.;
  \textsc{Albrecht}, M.; \textsc{Klimm}, D.; \textsc{Pietsch}, M.;
  \textsc{Brützam}, M.; \textsc{Bertram}, R.; \textsc{Ganschow}, S.;
  \textsc{Fornari}, R.:
\newblock {}\emph{Crystal Research Technologie} 45,  1229--1236, (2010).

\bibitem[11]{Cahill2}
\textsc{Cahill}, David~G.:
\newblock {}\emph{Review of Scientific Instruments} 61,  802--808, (1990).

\bibitem[12]{volklein}
\textsc{V{\"o}lklein}, F.; \textsc{Reith}, H.; \textsc{Cornelius}, T~W.;
  \textsc{Rauber}, M; \textsc{Neumann}, R:
\newblock {}\emph{Nanotechnology} 20,  325706, (2009).

\bibitem[13]{eindringtiefe}
\textsc{Ahmed}, Syed; \textsc{Liske}, Romy; \textsc{Wunderer}, Thomas;
  \textsc{Leonhardt}, Michael; \textsc{Ziervogel}, Ronny; \textsc{Fansler},
  Charlee; \textsc{Grotjohn}, Tim; \textsc{Asmussen}, Jes; \textsc{Schuelke},
  Thomas:
\newblock {}\emph{Diamond \& Related Materials} 15,  389--393, (2006).

\bibitem[14]{tritt}
\textsc{Tritt}, T.~M.:
\newblock \emph{Thermal Conductivity: Theory, Properties, and Applications}.
\newblock Springer, 2004

\bibitem[15]{kittel}
\textsc{Kittel}, Charles:
\newblock \emph{Introduction to Solid State Physics}.
\newblock John Wiley \& Sons, 2004

\bibitem[16]{Haying}
\textsc{He}, Haiying; \textsc{Blanco}, Miguel~A.; \textsc{Pandey}, Ravindra:
\newblock {}\emph{Applied Physics Letters} 88,  261904, (2006).

\bibitem[17]{silicon}
\textsc{Altukhov}, V.~I.; \textsc{Zavt}, G.~S.:
\newblock {}\emph{phys. stat. sol. (b)} 54, ~67, (1972).



\end{thebibliography}

\end{document}